\documentstyle{elsart}

\input epsf

\begin{document}
\begin{frontmatter}
\title{ Normal Mode Spectrum of the Deuteron in the Skyrme Model}

\author[Princeton]{C.\ Barnes}, 
\author[Swansea]{W.\ K.\ Baskerville\thanksref{PPARC}} and 
\author[Cambridge]{N.\ Turok} 
\address[Princeton]{Department of Physics, Princeton University, 
Princeton, \mbox{NJ 08544-0708} USA}
\address[Swansea]{Physics Department, University of Wales Swansea, 
Singleton Park, \mbox{Swansea SA2 8PP}}
\address[Cambridge]{Department of Applied Mathematics and Theoretical Physics, 
Silver Street, \mbox{Cambridge CB3 9EW} UK}
\thanks[PPARC]{Supported by PPARC}

\begin{abstract}
  The normal mode spectrum of the deuteron in the Skyrme model is
  computed. We find a bound doublet mode below the pion mass, which
  can be related to the well-known $90^{\circ}$ scattering of two
  skyrmions. We also find a singlet `breather' mode and another
  doublet above the pion mass. The qualitative pattern of the spectrum
  is similar to that recently found for the $B=4$
  multiskyrmion. The symmetries of all the vibrational modes are
  presented.
\end{abstract}
\end{frontmatter}

\section{Introduction}
\label{sec:intro}

The Skyrme model \cite{Skyrme1} has had some
success in describing both single nucleon properties and the
nucleon-nucleon interaction \cite{ANW,JackRho,Jacksons}. It is
therefore of interest to see how well the model performs for larger
nuclei. Classical multiskyrmion solutions are known up to baryon
number $B=9$ \cite{BC,BatSut}. However, these must be quantised
before any comparison to the real world can be made.

It is generally agreed that a proper treatment of the Skyrme model as
a quantum field theory is difficult.  Instead, a limited number 
of degrees of freedom are quantised, and usually
only the zero mode collective coordinates
\cite{ANW}. This 
reduces the model to finite-dimensional
quantum mechanics. The conventional wisdom is that $6N$ degrees of
freedom are required to describe a system containing $N$ nucleons (as
would be required for $N$ widely separated skyrmions). However,
minimal energy solutions for $B>1$ are in general single large
solitons, which have a maximum of 9 zero modes. Therefore even for
a very limited quantisation it is
necessary to include additional discrete modes.

This is the principal motivation for the calculation of the
vibrational spectra of multiskyrmions. A full quantisation should also
include the effects of the soliton on the vacuum fluctuations, or
radiation. We shall not attempt this here, but will merely present a
new method for computing normal mode spectra, with particular reference
to the $B=2$ solution. The results for $B=4$ are described in a
concurrent paper \cite{usB4}. A similar calculation was recently
performed by Walet \cite{Walet} for $B=2$ and $B=3$, using the
Yang-Mills instanton-induced form for the Skyrme fields. The main
differences in the present calculation is that we do not employ this
approximation, and also explicitly include a pion mass
term in the Lagrangian. This mass term clearly plays 
a key role in determining which vibrational modes are bound.

The vibrational spectra are interesting in their own right, especially
since we have discovered hitherto unsuspected similarities between the
two cases considered. In particular, the spectrum divides into two
types of modes. The lower modes correspond to those expected from an
approximate correspondence between BPS monopoles and Skyrmions
\cite{usB4}.  Going up in frequency, one encounters first the
`breather' and then higher multipole breathing modes. While the
structure of the spectrum is interesting, we emphasise that it
represents merely a first step in a rather more ambitious long term
project, namely the computation of properties of quantised
multiSkyrmions.

\section{Computation of the Spectrum}
\label{sec:method}

The basic idea behind our method for calculating the bound state
spectrum is quite simple. First, the classical solution is obtained by
numerical relaxation. It is then very slightly perturbed, and the
resulting configuration evolved for a long time at constant
energy. During this process, field values at certain points are
recorded as a function of time. This data can then be Fourier analysed
to obtain power spectra. Both the relaxation to the classical
solution, and the subsequent time evolution after perturbation,
require numerical solution of the Skyrme equations. The development of
a fast computer code to perform this task lies at the heart of our
method. In the current letter, we shall only briefly outline the main
ideas underlying our algorithm; further details may be found in
\cite{usbig}.

The Skyrme model has three free parameters: $F_{\pi}$, $e$ and
$m_{\pi}$. The work of Walet \cite{Walet} indicates that binding
energies after quantisation probably depend very sensitively on these
parameters. We defer further discussion of parameters
 until
the full calculation has been performed. Instead, we will express our
results in terms of the usual dimensionless Skyrme units
\cite{skyrme2}. In these units, the topological lower bound on the
energy is $12 \pi^{2} B$, and the Skyrme Lagrangian density is
\begin{equation}
  {\cal L} = \frac{1}{2}\, Tr(L_{\mu} L^{\mu}) + \frac{1}{16}\,
  Tr([L_{\mu}, L_{\nu}][L^{\mu}, L^{\nu}]) + 4 \beta^{2} (TrU -2),
  \label{Ldef}
\end{equation}
where $\beta = m_{\pi}/F_{\pi} e$ is a dimensionless constant,
$L_{\mu} = U^{\dagger} \partial_{\mu} U$ and $U = \sigma +
i\vec{\pi}.\vec{\tau}$ is the SU(2)-valued scalar field ($\tau_{i}$
are the Pauli matrices). For definiteness, we have set $\beta =
0.263$, following Adkins and Nappi\cite{AN}, although eventually we
would would hope to extend the calculation to treat $\beta$ as a free
parameter.

The action $S = \int \!\d^{4}\!x \,{\cal L}$ is discretised on a
finite lattice, and the equations of motion are derived by varying
with respect to each field component $\phi^{a}(i,j,k,t)$ (where $\phi
= (\sigma, \vec{\pi})$), at each point in the lattice. The $SU(2)$
constraint is enforced by the introduction of a term $\lambda
(\phi^{2} -1)$, where $\lambda$ is a Lagrange multiplier. At this
point we run into the usual problem that the ``kinetic'' (ie.\ time
dependent) part of the quartic Skyrme term causes coupling between
nearest neighbours in time and space. The problem is trivially solved
for the initial relaxation by simply ignoring the offending terms,
which are irrelevant for a static solution. But to compute the
normal modes we must consider
small fluctuations about the static solution:
$\phi = \phi_{0} + \delta \phi$. If we linearise in $ \delta \phi$
(just for the problematic terms), then the system of equations
uncouples, and we are left with an equation of the form
\begin{equation}
  \phi^{a}(i,j,k,t\!+\!1) = \tilde{\lambda}\, \phi^{a}(i,j,k,t) +
  R^{a}(i,j,k,t)
  \label{form}
\end{equation}
at each point, where $\tilde{\lambda}$ is a constant related to the
Lagrange multiplier. The $SU(2)$ constraint can then be invoked for
$\phi(t\!\!+\!\!1)$, to produce a quadratic equation which can be
solved for $\tilde{\lambda}$. This can then be substituted back into
Equation~(\ref{form}) to define the time evolution.

\section{Results and Interpretation of Spectra.}
\label{sec:results}

\begin{figure}[tb]
  \begin{center}
    \leavevmode
      {\hbox
      {\epsfxsize = 6.5cm \epsffile{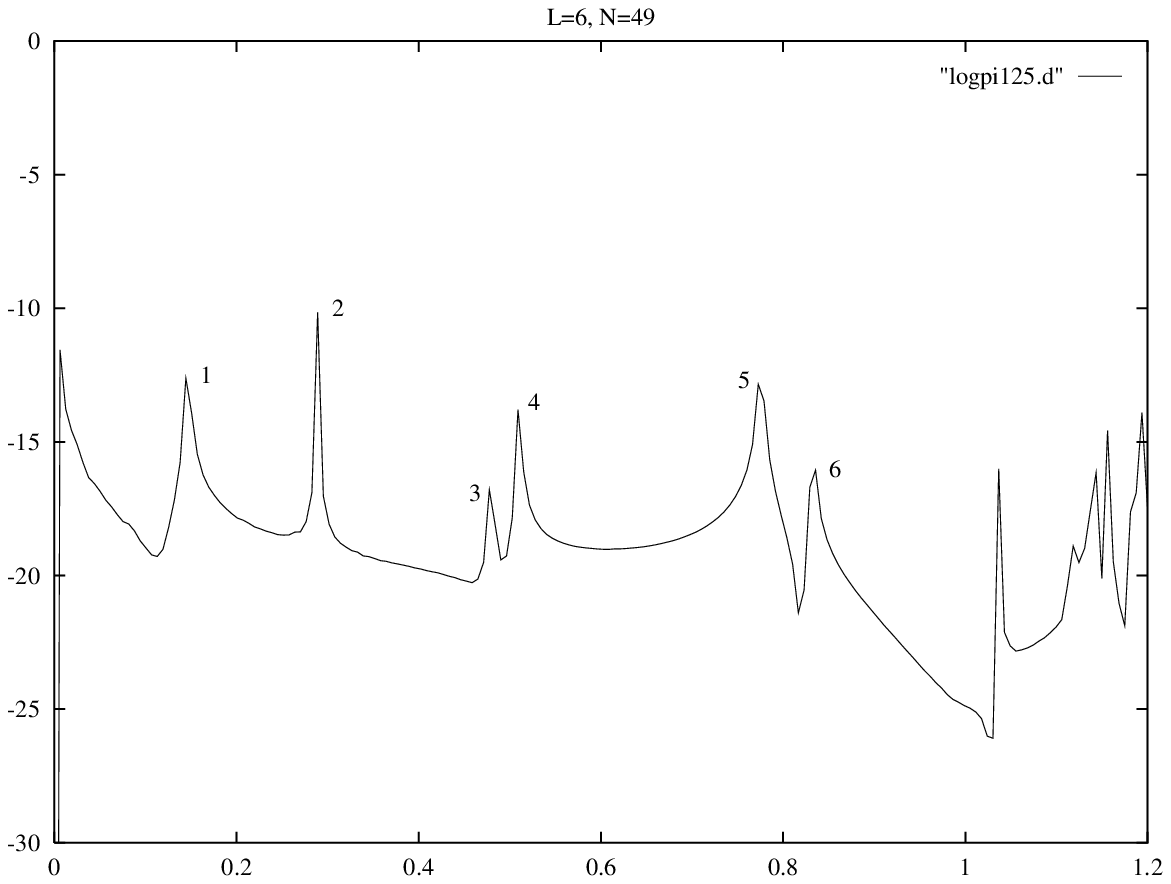}}
      {\epsfxsize = 6.5cm \epsffile{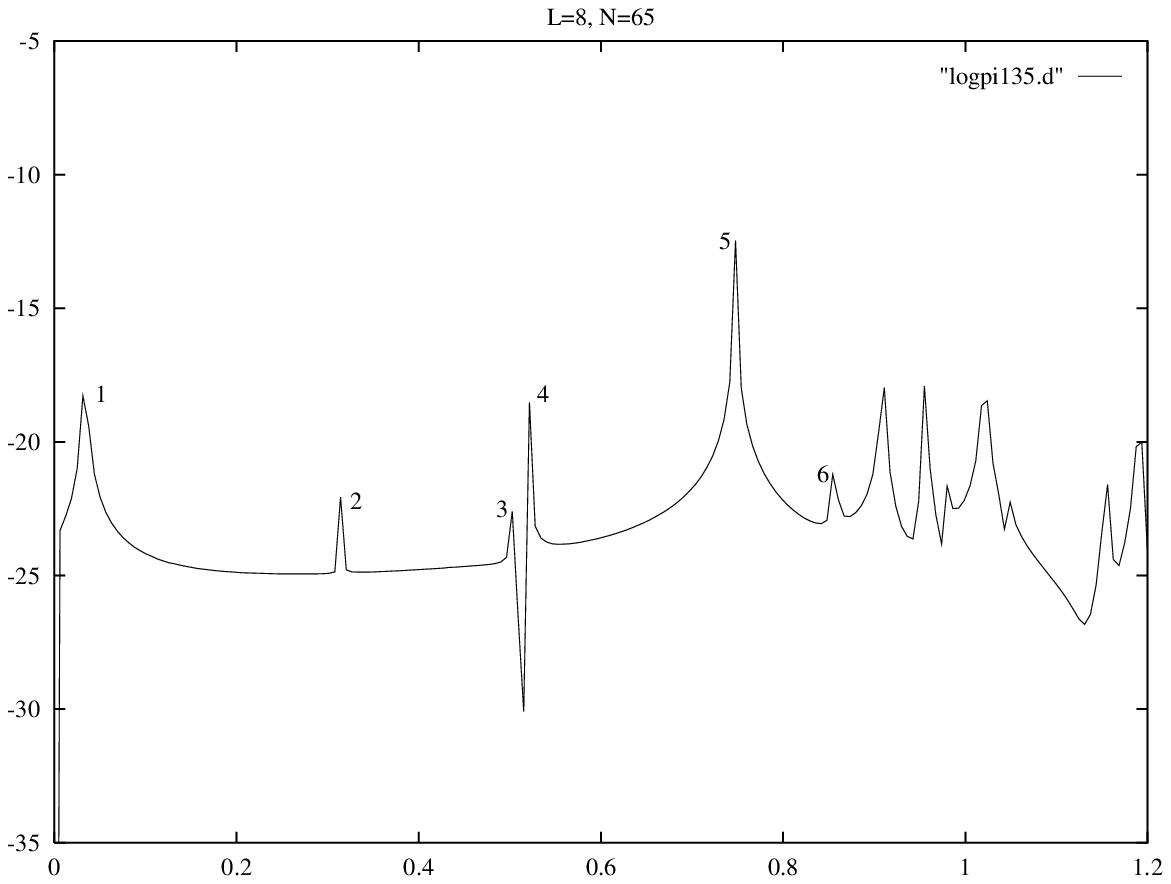}}
      }
  \end{center}
  \caption{Deuteron power spectra, for two different box sizes (L=6 and L=8). 
   Log(power) is plotted against angular frequency $\omega$.}
  \label{fig:power}
\end{figure}

Our simulations are carried out on a finite lattice of $N^{3}$ points,
of total volume $L^{3}$, so that the lattice spacing $\Delta x = L/N$.
Deuteron spectra have been calculated for a variety of box sizes and
latice spacings. We find that the frequencies observed show no
discernible dependence on lattice spacing, provided that 
the lattice
spacing is no larger than about 0.125 in our length units.
There is, however, significant dependence on box size.
Figure~\ref{fig:power} shows deuteron spectra for $L=6$ and $L=8$. Our
code uses periodic boundary conditions, so the finite
size effects can be understood in terms of the interaction between
neighbouring `image'  solitons. Since a pion mass term is included
explicitly in the Lagrangian, these effects should decay exponentially
as $L$ increases. The frequencies observed for $L=8$ should already be
a reasonable approximation to the infinite separation (box size)
limit; however, this can eventually be improved, with more data, by
fitting the exponential curve, and extrapolating.

Having obtained the normal mode spectrum, the next problem is to
interpret it. It is necessary to distinguish between vibrational modes
and radiation, and we also wish to determine the degeneracy and
symmetry of all modes. Note that radiative modes are discrete, rather
than forming a continuum above the pion mass, because of the finite
box size. For the same reason, bound states are possible above the
pion mass. In an infinite volume box, radiation continuum modes at the
frequencies of the vibrational modes would mix with the latter,
causing them to be spatially extended out to infinity.  In a finite
box, however, the relevant continuum modes may be absent, causing the
vibrational modes above the pion mass to be spatially localised.

To gain information about the observed peaks, the process of time
evolution is repeated, this time projecting out all previously
identified frequencies. We thus obtain Fourier amplitudes for
each mode, at all points on the lattice for each field component
\begin{equation}
  A^{a}_{\omega}(i,j,k) = \int\!\! \d t\: \delta
  \phi^{a}(i,j,k,t)\: cos \omega t 
  \label{Famp}
\end{equation}
where $\delta \phi^{a} = \phi^{a} - \phi_{0}^{a}$ is the difference
between the actual field value at a given time, and the static
solution. For each frequency, the total time of integration is chosen
to be an exact multiple of the period of oscillation.

The inner product between two such amplitudes can be defined
\begin{equation}
  \langle A_{\omega}|A_{\omega^{\prime}} \rangle =
  \int\!\! \d x^{3}\: K^{ab}(\vec{x})\: A^{a}_{\omega}(\vec{x})\:
  A^{b}_{\omega^{\prime}}(\vec{x}) ,
  \label{inner}
\end{equation}
with summation implied, where $K^{ab} = \delta^{ab}(1 + (\partial_{i}
\phi_{0})^{2}) - \partial_{i} \phi_{0}^{a} \partial_{j} \phi_{0}^{b}$
is the matrix which multiplies $\partial_{0}^{2} \phi^{b}$ on the left
hand side of the (linearised) Skyrme equations. It effectively plays
the role of a spatially dependent inertia tensor in the dynamics. The
inner product is defined in this way to ensure orthogonality between
modes of different frequency.

\begin{figure}[tb]
  \begin{center}
    \leavevmode
      {\hbox
      {\epsfxsize = 13cm \epsffile{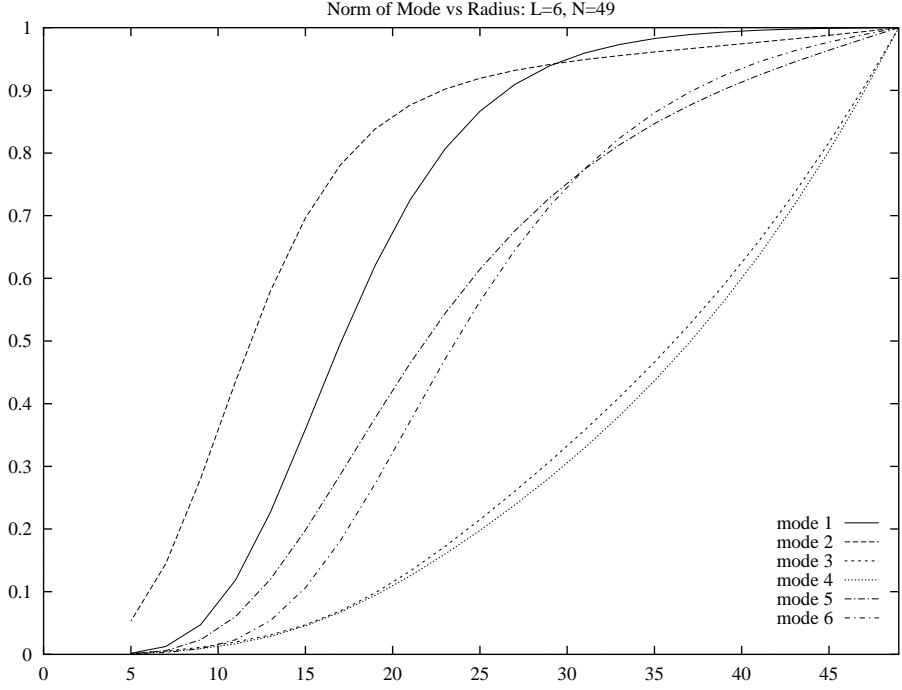}}
      }
  \end{center}
  \caption{$\langle A_{\omega}|A_{\omega} \rangle$ vs N, for the six
lowest modes in the deuteron spectrum. All modes are normalised so
that $\langle A_{\omega}|A_{\omega} \rangle =1$ when integrated over
the whole box}
  \label{fig:curves}
\end{figure}

The most immediate concern is to distinguish between bound modes and
radiation in the spectrum. To do this, it is useful to calculate the
norm of a given mode as a function of radius. Instead of integrating
$\langle A_{\omega}|A_{\omega} \rangle$ over the whole box, we instead
integrate only for $N^{3}$ points in a cube in the middle. By varying
$N$, $\langle A_{\omega}|A_{\omega} \rangle$ is mapped out as a
function of ``radius'' (or side length). If a particular mode is
bound, then it should be localised at or near the static
solution. Thus the curve of $\langle A_{\omega}|A_{\omega} \rangle$ vs
$N$ should rise steeply at a radius corresponding roughly to the
soliton, then flatten out as the edge of the box is
approached. Radiation, on the other hand, should be spread evenly
throughout the box, giving a curve going approximately as $N^{3}$,
rising most steeply at the edge. The curves for the six modes labelled
in Figure~\ref{fig:power} are plotted in Figure~\ref{fig:curves}. It
is clear that the curves do indeed display characteristic shapes, so
that this method successfully provides a clear distinction between
bound modes and radiation. We conclude that modes 3 and 4 are
radiative, whereas modes 1, 2, 5 and 6 are bound.

The next obvious question is the degeneracy of all the modes. If a
given frequency $\omega$ is $n$-fold degenerate, then an $m \times m$
matrix of inner products $\langle A_{\omega}^{i}|A_{\omega}^{j}
\rangle$ (where $m > n$), will only have $n$ non-zero eigenvalues.
Fourier amplitudes from different runs, with different random initial
perturbations, (but for the same frequency), can be used to calculate
the degeneracies via this method. However, cleaner results are often
obtained by taking one amplitude, then generating more by applying
different symmetry operations. The symmetries used must leave the
static solution invariant; they generally consist of a physical
transformation followed by an isospin transformation of the pion
fields. Using this method, we find that modes 3 and 5 are singlets,
while peaks 1, 2, 4, and 6 correspond to doublets.

The only remaining issue concerns symmetry. We wish to classify all
the modes according to the representation they form of the symmetry
group of the static solution. The classical $B=2$ multiskyrmion is
axially symmetric, and in addition possesses a reflection symmetry in
the plane of the torus. The symmetry group is $D_{\infty h}$, axial
symmetry extended by inversion. Its character table is given in
Table~\ref{tab:char}. The notation of Hamermesh\cite{Ham} has been
used for the conjugacy classes and the singlet representations. $E$ is
the identity, $C(\phi)$ is rotation by angle $\phi$, $\sigma_{v}$ is
reflection in any plane including the $z$-axis, and $I$ is
inversion. The singlet representations are labelled $g$ or $u$ to
indicate whether they are even or odd under parity, while the $+$ or
$-$ refers to their behaviour under $\sigma_{v}$. There are an
infinite number of doublets, labelled by an integer $n$, and parity.
\begin{table}[b]
  \begin{center}
    \leavevmode
    \begin{tabular}{|c|c|c|c|c|c|c|} \hline
        \  & $E$ & $C(\phi)$ & $\sigma_{v}$ & $I$ &
        $I C(\phi)$ & $I \sigma_{v}$ \\ \hline
        $\Sigma^{+}_{g}$ & 1 & 1 & 1 & 1 & 1 & 1 \\
        $\Sigma^{+}_{u}$ & 1 & 1 & 1 & -1 & -1 & -1 \\
        $\Sigma^{-}_{g}$ & 1 & 1 & -1 & 1 & 1 & -1 \\
        $\Sigma^{-}_{u}$ & 1 & 1 & -1 & -1 & -1 & 1 \\
        $n^{+}$ & 2 & $2 \cos n\phi$ & 0 & 2 & $2 \cos n\phi$ & 0 \\
        $n^{-}$ & 2 & $2 \cos n\phi$ & 0 & -2 & $2 \cos n\phi$ & 0 \\ \hline
    \end{tabular}
  \end{center}
  \caption{Character table for $D_{\infty h}$}
  \label{tab:char}
\end{table}

It is clear that singlet representations are distinguished by their
properties under $\sigma_{v}$ and $I$. The symmetries of modes 3 and 5
can therefore be determined by taking matrix elements of these
operators
\begin{eqnarray}
  \langle 3|\sigma_{v}|3 \rangle = 1 & \ \ \ \ \ \ \ \ &
  \langle 3|I|3 \rangle = -1 \nonumber \\
  \langle 5|\sigma_{v}|5 \rangle = 1 & \ \ \ \ \ \ \ \ &
  \langle 5|I|5 \rangle = 1 .
  \label{singlet}
\end{eqnarray}

For doublet states, the simplest way to determine $n$ is to search for
the lowest angle $\phi$ such that $\langle X|C(\phi)|X \rangle =0$
(this implies that $\langle X|C(2\phi)|X \rangle =-1$ and $\langle
X|C(4\phi)|X \rangle =1$). If the lowest angle $\phi = 90^{\circ}$
then $n=1$, if $\phi = 45^{\circ}$ then $n=2$, and so on. In practice,
$n=1$ or 2 are the only realistic values, so the matrix elements of
$C(45^{\circ})$ and $C(90^{\circ})$, together with inversion, provide
sufficient information to distinguish the representations.
\begin{equation}
\begin{array}{lll}
  \langle 1|C(45^{\circ})|1 \rangle = \frac{1}{\sqrt{2}} \ \ \ \ \ \ \ \ & 
  \langle 1|C(90^{\circ})|1 \rangle = 0 \ \ \ \ \ \ \ \ &
  \langle 1|I|1 \rangle = 1 \nonumber \\
  \langle 2|C(45^{\circ})|2 \rangle = 0 \ \ \ \ \ \ \ \ &
  \langle 2|C(90^{\circ})|2 \rangle = -1 \ \ \ \ \ \ \ \ &
  \langle 2|I|2 \rangle = 1 \nonumber \\
  \langle 4|C(45^{\circ})|4 \rangle = 0 \ \ \ \ \ \ \ \ &
  \langle 4|C(90^{\circ})|4 \rangle = -1 \ \ \ \ \ \ \ \ &
  \langle 4|I|4 \rangle = 1 \nonumber \\
  \langle 6|C(45^{\circ})|6 \rangle = \frac{1}{\sqrt{2}} \ \ \ \ \ \ \ \ &
  \langle 6|C(90^{\circ})|6 \rangle = 0 \ \ \ \ \ \ \ \ &
  \langle 6|I|6 \rangle = -1 \nonumber \\
  \label{doublet}
\end{array}
\end{equation}

To summarise, the first peak is a doublet, with symmetry
$1^{+}$. This is the representation under which axial vectors
transform in $D_{\infty h}$. These states do not correspond to true
vibrations, but rather represent two broken zero modes, spatial
rotations of the torus around the $x$ and $y$ axes. The periodic
boundary conditions of the lattice introduce the possibility of low
energy `spin waves', whose frequency will decrease as the box size
increases, eventually reverting to zero modes in the infinite limit.
This explains the extreme shift in frequency observed for this mode
when the box size is changed from $L=6$ to $L=8$.

The second peak in the spectrum represents the first true finite
frequency bound modes. They are doubly degenerate, with a quadrupole
symmetry $2^{+}$. If amplified, these would become the classic
$90^{\circ}$ scattering of two skyrmions. There is also a close
analogy to the scattering of BPS monopoles.

The zero modes of the two-monopole toroidal BPS solution may be fairly
simply understood. The BPS solutions are associated with rational maps
constructed as follows. A complex variable $z$ is associated with each
direction in space through a projective map. The complex $z$ plane is
embedded in three Euclidean dimensions, and a unit two-sphere is drawn
centered on the origin.  Each direction $\vec{n}$ defines a point on
the two-sphere, which is then projected from one of the poles of the
two-sphere onto the $z$ plane. One then defines a rational map $R$
which is a ratio of polynomials in $z$, each up to $z^N$ for a
N-monopole solution. Just as $z$ defines a direction in real space,
$R$ defines one in internal space, acted on by the global $SU(2)$
symmetry via Moebius transformations $R \rightarrow (\alpha R+\beta)/
(-\beta^*R+\alpha^*)$, with $\alpha$ and $\beta$ being the 11 and 12
matrix elements of the $SU(2)$ matrix.

For special choices of $R$, the solutions exhibit enlarged symmetry.
In our case, the 2-monopole solution represented by $R=z^2$ is the
toriodal solution which has $D_{\infty h}$ symmetry.  It is invariant
under rotations $z \rightarrow e^{i\phi} z$ combined with isospin
rotations $R \rightarrow e^{-2i\phi} R$, and also inversions $z
\rightarrow -1/z^*$, $R \rightarrow 1/R^*$. The general perturbed
2-monopole solution is represented by $R=(z^2+az+b)/(cz^2+dz+1+e)$
where the coefficient of $z^2$ in the numerator is taken to be unity
and the complex parameters $a$, $b$, $c$, $d$, $e$ are small. These
parameters are not invariant under $D_{\infty h}$, but transform
according to some irreducible representation, which has ten real
dimensions.  It is straightforward to check by computing the
characters of the symmetry group elements that the representation is
$2^+$ $+$ $2^-$+ $1^+$ $+$ $\Sigma^{-}_{g}$ $+$ $1^-$ $+$
$\Sigma^{-}_{u}$.  The isospin zero modes comprise the $2^-$
representation, the rotational zero modes comprise the $1^+$ $+$
$\Sigma^{-}_{g}$ representation and the tranlational zero modes the
$1^-$ $+$ $\Sigma^{-}_{u}$ representation.  This leaves the $2^+$
representation as the remaining nontrivial zero mode of the 2-monopole
BPS solution.  And this representation coincides with our results for
the lowest vibrational mode of the deuteron in the Skyrme model.

The next two frequencies correspond to radiation, with wave number
$k=0$. These modes are split, rather than forming a degenerate
triplet, because the pion fields of the static solution do not all
transform the same way. The radiative modes fall into the same
representations as the static solution pions: $\Sigma_{u}^{+} +
2^{+}$.

The fifth peak is a breathing mode, where the size of the soliton
fluctuates with no change in shape. It is a singlet, and transforms
trivially under $D_{\infty h}$, as would be expected. The sixth peak
represents the last bound states. These form a doublet, with symmetry
$1^{-}$. The torus expands on one side while contracting on the other,
a dipole `breathing' motion. Altogether then, we have found five
finite energy normal modes in the deuteron spectrum.

\section{Conclusions.}
\label{sec:conclusions}

The normal mode spectrum of the deuteron has been calculated. The
nature of all states has been identified, and their degeneracies and
symmetries calculated. We find a bound doublet below the pion mass,
which is related to $90^{\circ}$ scattering between two skyrmions. Two
more bound states appear above the pion mass: a breather, plus a
second doublet with a dipole symmetry.

The overall pattern of the deuteron spectrum displays a remarkable
similarity to that recently found for the $B=4$ solution \cite{usB4}.
There again, all modes below the pion mass can be related to those
expected from an approximate correspondence with the zero modes of BPS
multimonopoles. The first $B=4$ bound mode above the pion mass is a
breather, and the remaining bound modes involve `breathing' motion of
the baryon density. It is tempting to speculate that a similar pattern
will be observed for all multiskyrmion spectra, but more examples are
required to confirm this.

One further remark may be worthwhile. We have found five finite energy
bound modes in the deuteron spectrum. Added to the eight zero modes,
this makes a total of 13. Sixteen bound modes were found for $B=4$,
giving a total of 25. In both cases, then, there is a total of $6B +1$
modes, one more than is usually expected. The ``extra'' mode is the
breather; every multiskyrmion, including the single skyrmion, must
have such a mode. The remaining vibrational modes can be considered as
corresponding to the broken zero modes of the separated skyrmions.

\begin{figure}[tb]
  \begin{center}
    \leavevmode
        {\hbox
        {\epsfxsize = 6.5cm \epsffile{one.eps}}
        {\epsfxsize = 6.5cm \epsffile{two.eps}}
        }
        {\hbox
        {\epsfxsize = 5.5cm \epsffile{three.eps}}
        {\epsfxsize = 6.5cm \epsffile{four.eps}}
        }
        {\hbox
        {\epsfxsize = 5.8cm \epsffile{five.eps}}
        {\epsfxsize = 6.5cm \epsffile{six.eps}}
        }

  \end{center}
  \caption{Deuteron vibrational modes: plots of ${\cal B}(\phi_{0} + \delta
    \phi) - {\cal B}(\phi_{0})$. Isosurfaces indicate surfaces of
    constant negative perturbation; contours positive. Note that all
    modes except mode 3 are plotted looking more or less directly down
    the $z$-axis, so that whereas modes 2 and 4 are quadrupoles in the
    $(x,y)$-plane, mode 1 is a dipole, but is odd under reflections in
    this plane. Mode 3 is odd under $z$-reflections, but the twisting
    is merely an artifact due to slight contamination from mode 2.}
  \label{fig:pictures}
\end{figure}

\begin{ack}
  We would like to thank N. Manton and C. Houghton for useful
  discussions of this work.
\end{ack}


\begin{thebibliography}{99}
\bibitem{Skyrme1} 
T. H. R. Skyrme, {\em Proc. Roy. Soc.\/} {\bf 260} (1961) 127 
\bibitem{ANW} 
G.S. Adkins, C.R. Nappi and E. Witten, {\em Nucl. Phys. \/} {\bf B228}
(1983) 552  
\bibitem{JackRho} 
A.D. Jackson and M. Rho, {\em Phys. Rev. Lett. \/} {\bf 51} (1983) 751
\bibitem{Jacksons}
A. Jackson and A.D. Jackson, {\em Nucl. Phys. \/} {\bf A457} (1986)
687 
\bibitem{BC}
E. Braaten, S. Townsend and L. Carson, {\em Phys. Lett. \/} {\bf B235} (1990)
147 
\bibitem{BatSut}
R.A. Battye and P.M. Sutcliffe, preprint UKC-IMS-97-07,
IMPERIAL-TP-96-97-18 
\bibitem{Walet}
N. Walet, {\em Nucl. Phys. \/} {\bf A606} (1996)
429 
\bibitem{AN} 
G.S. Adkins and C.R. Nappi, {\em Nucl. Phys. \/} {\bf B233}
(1984) 109  
\bibitem{usB4}
C. Barnes, K. Baskerville and N. Turok, `Bound State Spectrum of the
$B=4$ Multiskyrmion', hep-th/9704012, {\em Phys. Rev. Lett.\/},
submitted 
\bibitem{usbig}
in preparation
\bibitem{Ham}M. Hamermesh, {\it Group Theory and its Application to Physical Problems},
Dover, 1962.
\bibitem{skyrme2}
T.H.R. Skyrme, {\em Nucl. Phys. \/} {\bf 31} (1962) 556

\end{thebibliography}
\end{document}